\documentclass{TEMA}

\usepackage[brazil]{babel}      

\usepackage[latin1]{inputenc}   
\usepackage{subfigure}
\usepackage{graphicx}

\newcommand{\pt}{\partial}

\begin{document}

\title
    {Simulação Numérica da Dinâmica de Coliformes Fecais no Lago Luruaco, Colômbia \thanks{Trabalho apresentado no XXXVI Congresso Nacional de Matemática Aplicada e Computacional.}}

\author
    {T.M. SAITA%
     \thanks{tatianasaita@gmail.com} \,, P.L. NATTI \thanks{plnatti@uel.br}\,,
     E.R. CIRILO \thanks{ercirilo@uel.br} \,,
     N.M.L. ROMEIRO \thanks{nromeiro@uel.br} \,,
     Departamento de Matemática, CCE, Universidade
     Estadual de Londrina, Rodovia Celso Garcia
     Cid, 900 Campus Universitário, 86057-970, Londrina, PR, Brasil.\\ \\
     M.A.C. CANDEZANO \thanks{mcaro72@gmail.com} \,, 
     Departamento de Matemática, Universidad del Atlántico,  
     Barranquilla, Atlántico, Colômbia.\\ \\
     R.B ACU\~NA \thanks{rafaelborja@mail.uniatlantico.edu.co} \,, 
     L.C.G. MORENO, \thanks{luiscarlosgutierrez@mail.uniatlantico.edu.co} \,, 
     Departamento de Biología, Universidad del Atlántico,  
     Barranquilla, Atlántico, Colômbia}

\criartitulo
\runningheads {SAITA, NATTI, CIRILO, ROMEIRO, CANDEZANO, AC\~UNA e MORENO}{Simulação numérica da dinâmica de coliformes fecais no lago Luruaco, Colômbia}

\begin{abstract}
{\bf Resumo}

O lago Luruaco localizado no Departamento de Atlántico, Colômbia, sofre com o despejo de esgoto sem tratamento, trazendo riscos à saúde de todos 
que utilizam suas águas. O presente estudo tem como objetivo realizar a simulação numérica da dinâmica da concentração de coliformes fecais no 
lago. A simulação do fluxo hidrodinâmico do lago é realizada por meio de um modelo bidimensional horizontal (2DH), dado por um sistema de 
equações de Navier-Stokes. Já a simulação do transporte de coliformes fecais é descrita por uma equação convectiva-dispersiva-reativa. Essas 
equações são resolvidas numericamente pelo Método de Diferenças Finitas (MDF) e pelo método Mark and Cell (MAC), em coordenadas generalizadas. 
Quanto à construção da malha computacional do lago Luruaco, os métodos spline cúbica e multibloco foram utilizados. Os resultados obtidos nas 
simulações permitiram uma melhor compreensão da dinâmica de coliformes fecais no lago Luruaco, evidenciando as regiões mais poluídas. Os resultados
também podem orientar órgãos públicos quanto à identificação dos emissores de poluentes no lago e o desenvolvimento de um tratamento otimizado 
para a recuperação do ambiente poluído.

{\bf Palavras-chave}. Lago Luruaco, Coliformes Fecais, Malha Multibloco, Método de Diferenças Finitas, Método MAC, Coordenadas Generalizadas.
\end{abstract}

\newsec{Introdução}

A poluição hídrica é um problema mundial que prejudica a saúde, a segurança e o bem-estar da população, afetando desfavoravelmente todos os seres 
vivos de um determinado ambiente. Um caso particular de poluição hídrica é a contaminação por um produto ou organismo, tal como o lançamento de
esgotos domésticos em corpos d'água. Como fatores negativos da poluição hídrica tem-se a deterioração da qualidade da água, a proliferação de 
doenças, a morte de espécies aquáticas, a eutrofização, entre outros. Segundo a Organização Mundial da Saúde (OMS) pelo menos 2 milhões de 
pessoas, principalmente crianças com menos de 5 anos de idade, morrem por ano no mundo devido a doenças causadas pela água contaminada 
\cite{OMSpoluicao}. 

O corpo d'água em estudo, o lago Luruaco localizado no Departamento de Atlántico, Colômbia, sofre com o problema da poluição, principalmente 
devido ao esgoto gerado por cerca de 28000 habitantes do município de Luruaco, cidade situada à montante do lago \cite{herreraplan}. As consequências da
contaminação do lago Luruaco são sentidas pelos próprios moradores da região, que utilizam essa água para o consumo diário. 

A análise periódica da qualidade da água é uma importante ferramenta de auxílio para que os órgãos responsáveis possam avaliar os riscos que uma poluição pode causar à população. A OMS recomenda que as bactérias do grupo coliformes fecais sejam
utilizadas como parâmetro microbiológico de qualidade d'água, quando se deseja mensurar a presença de organismos patogênicos \cite{world1987guias}. Nos últimos anos um grande esforço tem sido realizado para desenvolver modelos matemáticos que descrevam adequadamente a dinâmica de coliformes fecais em diferentes corpos d'água 
\cite{Servais,LIU,romeiro}. Nesse contexto, estudar a dinâmica de coliformes fecais no corpo d'água do lago Luruaco é uma forma de determinar as regiões que apresentam maior risco de contaminação para a população local.

A utilização de modelos matemáticos para a análise dos parâmetros de qualidade da água começou a se desenvolver em 1925, a partir do modelo 
unidimensional de Streeter-Phelps que avaliava a demanda bioquímica de oxigênio (DBO) e o oxigênio dissolvido (OD) \cite{streeter}. Com o 
desenvolvimento da tecnologia e com a maior necessidade de estudar problemas de poluição ambiental, outros modelos matemáticos foram 
desenvolvidos e se tornaram cada vez mais complexos, descrevendo também propriedades biológicas ou químicas do corpo d'água \cite{couto,salvage}. 
Dentre os principais modelos desenvolvidos citam-se HSPF, MIKE, QUAL2E, QUASAR, SIMCAT e WASP \cite{Sharma,Tsaris}.
Embora esses modelos já sejam bem testados e desenvolvidos, eles são modelos comerciais e do tipo ''caixa preta''. Em geral, eles não 
permitem que as equações e/ou modelo numérico sejam alterados conforme as necessidades do pesquisador. Neste trabalho, ao invés de 
utilizar modelos tipo ''caixa preta'', optou-se por modelar a dinâmica de concentrações de coliformes fecais no lago Luruaco a partir 
das equações de conservação, e implementar um modelo numérico compatível com o problema. Assim, caso seja necessário, é possível 
modificar: (a) o modelo matemático, agregando novos termos, (b) o modelo geométrico, através de readequações via coordenadas 
generalizadas, (c) o modelo numérico, implementando novos esquemas numéricos que tratam termos convectivos.

O objetivo desse trabalho é definir um modelo representativo da dinâmica de coliformes fecais no lago Luruaco, auxiliando assim no monitoramento da qualidade das águas do referido corpo hídrico. Neste contexto, utiliza-se as bactérias do 
tipo coliformes fecais (presentes principalmente no esgoto doméstico) como parâmetro para a qualidade da água. A dinâmica de concentração dos 
coliformes fecais é obtida por meio da simulação numérica de um modelo bidimensional horizontal (2DH) durante um determinado 
intervalo de tempo \cite{romeiro}. O modelo de transporte de coliformes fecais é dado por uma equação convectiva-dispersiva-reativa, cujo termo
convectivo contém o campo de velocidades do fluxo hidrodinâmico, obtido a
partir de um sistema de equações de Navier-Stokes. 

As técnicas de discretização mais utilizadas em análise numérica são o Método de Diferenças Finitas (MDF), o Método de Volumes Finitos (MVF) 
e o Método de Elementos Finitos (MEF) \cite{ferziger}. Em nosso estudo a discretização dessas equações diferenciais é realizada pelo 
Método de Diferenças
Finitas devido à simplicidade matemática e a fácil implementação computacional. Por se tratar de um modelo não linear são adotados 
métodos específicos 
para a discretização desses termos que contém não 
linearidades \cite{romeiro}. O esquema \textit{First Order Upwind} (FOU)  é aplicado e a solução do sistema hidrodinâmico é obtida 
utilizando o método \textit{Mark and Cell} (\textit{MAC}), gerando o campo de velocidades para as águas do lago Luruaco.      

Quanto ao domínio físico do problema (lago), ele é construído com o uso do método multibloco em coordenadas generalizadas, onde a malha 
computacional é feita a partir da 
união de sub-malhas menores. Uma das vantagens desse método é a de preservar as características da geometria e mapear com mais fidelidade  
a região de interesse, com a possibilidade de 
refinar apenas regiões específicas da malha. Atualmente esse método é bastante utilizado em aplicações de problemas realísticos \cite{king}.

O artigo é apresentado na seguinte forma: na Seção 2 apresenta-se as propriedades físicas e geográficas do lago Luruaco. Na Seção 3 
descreve-se a Modelagem Matemática e Numérica desenvolvida para esse estudo. Na Seção 4 a Modelagem Geométrica do interior da malha do 
lago é realizada através do método multiblocos, enquanto o contorno é ajustado por splines cúbicas. Enfim, na Seção 5 são apresentadas 
e discutidas as simulações numéricas da dinâmica de coliformes fecais no lago Luruaco. Considerações finais também são apresentadas.

\newsec{O Lago Luruaco}

O lago Luruaco está localizado no departamento de Atlántico, Colômbia, entre as coordenadas 10º16' e 11º04' de latitude norte e 74º43' e 75º16' de 
longitude oeste. O lago está a 31 metros de altitude em relação ao nível do mar, ocupando uma região de cerca de 420 hectares com profundidade 
média de 5 metros. Estima-se que a capacidade de armazenamento de água no lago seja de 12,5 $\times$ 10$^6$ m$^3$ \cite{diagnostico}.

Situado à noroeste da Colômbia, o lago Luruaco está numa região de clima tropical, apresentando características predominantemente quentes durante 
o ano com temperaturas variando de 24ºC a 28ºC. 
Os ventos na região sopram no sentido nordeste-sudeste com velocidade média de 15 a 20 km/h. 
O regime de precipitações, quando não há a influência de fenômenos climáticos (El Niño e La Niña), ocorre nos períodos entre maio/junho e agosto/novembro, alternando 
com períodos secos. Por outro lado, nos anos que tais fenômenos são mais intensos, longos períodos de estiagem (devido ao El Niño) ou de chuvas intensas (devido ao El Niña) ocorrem  \cite{diagnostico}. 

O lago Luruaco pertence à bacia do Canal del Dique e depende de riachos para ser abastecido. Os principais riachos ligados ao lago Luruaco são: riacho Limón, riacho Mateo e riacho Negro. O córrego que comunica o lago Luruaco ao lago San Juan de Tocagua faz o papel de efluente do lago, devido à diferença de altitude entre os lagos. A Figura \ref{fig1} indica a localização dos afluentes e efluente no lago Luruaco.

\begin{figure}[!ht]
		\centering
		\includegraphics[scale=0.39]{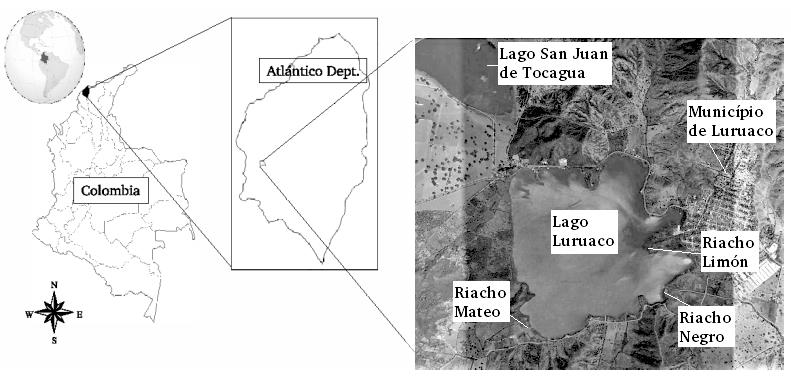}
	\caption{Localização das entradas de água do lago Luruaco através do Riacho Mateo, Riacho Negro e Riacho Limón e saída através do 
	canal para o lago San Juan de Tocagua. Fonte: Adaptado de Google Earth, 2016.}
	\label{fig1}
\end{figure}

A água do lago é utilizada em atividades como agricultura, pecuária, pesca, turismo e consumo. Além disso, quase às margens do lago, está situado 
o Município de Luruaco, que possui cerca de 28000 habitantes \cite{herreraplan}. Essa população depende diretamente do lago. 
Por outro lado, o esgoto doméstico gerado pelos moradores do município de Luruaco e regiões próximas são uma fonte de matéria orgânica 
para o lago. Sem o tratamento adequado, a maior parte do esgoto doméstico gerado pelas cerca de 5000 habitações \cite{diagnostico}, são 
introduzidos 
no meio ambiente através de fontes difusas, isto é, não possuem pontos de lançamentos específicos. Somado a isso, existe também, os 
resíduos da agricultura e pecuária. Esta contaminação é canalizada para os riachos que abastecem o lago Luruaco, alterando a composição 
física e química da água, e em algumas situações tornando-a imprópria para o consumo \cite{diagnostico}.

\newsec{Modelo Matemático e Numérico}

O modelo utilizado neste estudo fornece a concentração de coliformes fecais na superfície do lago Luruaco através de um sistema de equações  diferenciais 
parciais bidimensionais horizontais (2DH). 
O lago Luruaco não tem alta concentração de coliformes fecais em suspensão no corpo d'água, de modo que nossa suposição é que os coliformes fluem com o campo de 
velocidades hidrodinâmico do lago. Nessa situação diz-se que os coliformes fecais estão em regime passivo e o estudo de seu transporte pode ser realizado de modo 
desacoplado do modelo hidrodinâmico. Neste contexto, nossa modelagem matemática supõem que os coliformes fecais escoam com a mesma velocidade 
do fluido \cite{chapra, Suellen}.

Nesse trabalho o modelo hidrodinâmico das águas do lago Luruaco é dado por um sistema de equações de Navier-Stokes, onde 
considera-se a água incompressível de viscosidade constante, isto é, $\rho=cte$ e $\mu=cte$. Esse sistema, em conjunto com a
equação da continuidade, em coordenadas generalizadas $(\xi, \eta, \tau)$, escreve-se como:

\begin{eqnarray} 
\underbrace{\frac{\pt U}{\pt \xi}+ \frac{\pt V}{ \pt \eta}}_{\mbox{{\small termo da eq. de continuidade}}} &=&0 \; \label{eq1} \\
\nonumber \\ 
\nonumber \\
\underbrace{\frac{\pt}{\pt \tau} \left(\frac{ u}{J} \right)}_{\mbox{{\small termo temporal}}} +\underbrace{\frac{\pt}{\pt \xi} ( U u ) +\frac{\pt}{\pt \eta} ( V u)}_{\mbox{{\small termo convectivo}}} & = & \frac{1}{\rho} \underbrace{\left(\frac{\pt p}{\pt \eta} \frac{\pt y}{\pt \xi}-\frac{\pt p}{\pt \xi} \frac{\pt y}{\pt \eta} \right)}_{\mbox{{\small termo de pressão}}} + \nonumber \\
\nu \underbrace{\left[\frac{\pt}{\pt \xi} \left( J \left( \alpha \frac{\pt u}{\pt \xi} -\beta \frac{\pt u}{\pt \eta} \right)\right)\right.}_{\mbox{{\small termo difusivo}}}
& + & \underbrace{ \left. \frac{\pt }{\pt \eta} \left(J \left( \gamma \frac{\pt u}{\pt \eta} - \beta \frac{\pt u}{\pt \xi} \right) \right) \right]}_{\mbox{{\small termo difusivo}}} \label{eq2}\\
\nonumber \\
\nonumber \\
\underbrace{\frac{\pt}{\pt \tau} \left(\frac{ v}{J} \right)}_{\mbox{{\small termo temporal}}} + \underbrace{\frac{\pt}{\pt \xi} ( U v ) +\frac{\pt}{\pt \eta} ( V v)}_{\mbox{{\small termo convectivo}}} & = & \underbrace{\frac{1}{\rho} \left(\frac{\pt p}{\pt \xi} \frac{\pt x}{\pt \eta} -\frac{\pt p}{\pt \eta} \frac{\pt x}{\pt \xi} \right)}_{\mbox{{\small termo de pressão}}} + \nonumber \\
\nu \underbrace{\left[\frac{\pt}{\pt \xi} \left( J \left( \alpha \frac{\pt v}{\pt \xi} -\beta \frac{\pt v}{\pt \eta} \right)\right) \right.}_{\mbox{{\small termo difusivo}}}
& + &  \underbrace{\left. \frac{\pt }{\pt \eta} \left(J \left( \gamma \frac{\pt v}{\pt \eta} - \beta \frac{\pt v}{\pt \xi} \right) \right) \right]}_{\mbox{{\small termo difusivo}}} \label{eq3}
\end{eqnarray}

\noindent onde $(u, v)$ é o campo de velocidades do escoamento, $U$ e $V$ são as componentes contravariantes do campo de 
velocidades, $\rho$ e $\nu=\mu / \rho$ são a densidade e a viscosidade cinemática constantes, respectivamente, $p$ é a pressão, $J$ é o 
Jacobiano \cite{maliska} dado por 
\begin{equation}
J= \left[ \frac{\pt x}{\pt \xi}\frac{\pt y}{\pt \eta}- \frac{\pt x}{\pt \eta}\frac{\pt y}{\pt \xi}\right]^{-1}
\label{eq5}
\end{equation}
 
 \noindent e as quantidades $\alpha$, $\beta$ e $\gamma$ são dadas por 
 \begin{equation}
 	\alpha= \left[ \frac{\pt x}{\pt \eta}\right]^2 + \left[ \frac{\pt y}{\pt \eta}\right]^2 \;\;,\;\;
 	\beta=  \frac{\pt x}{\pt \xi}\frac{\pt x}{\pt \eta} + \frac{\pt y}{\pt \xi}\frac{\pt y}{\pt \eta} \;\;,\;\;
 	\gamma= \left[ \frac{\pt x}{\pt \xi}\right]^2 + \left[ \frac{\pt y}{\pt \xi}\right]^2 . \nonumber
 	\label{eq5a}
 \end{equation}
 
 \noindent O sistema (\ref{eq1}-\ref{eq3}) fornece o campo hidrodinâmico de velocidades da água no lago Luruaco. O desenvolvimento deste modelo 
 pode ser encontrado em \cite{alessandra}.
 
Do campo de velocidades da água e do modelo de transporte pode-se determinar a dinâmica de concentração de coliformes fecais em todo lago 
Luruaco. Nesse trabalho o modelo de transporte convectivo-dispersivo-reativo em coordenadas generalizadas é dado por

\begin{eqnarray}
\underbrace{\frac{\pt}{\pt \tau} \left(\frac{C}{J} \right)}_{\mbox{{\small termo temporal}}}+\underbrace{\frac{\pt}{\pt \xi} (UC ) +\frac{\pt}{\pt \eta} (VC)}_{\mbox{{\small termo convectivo}}}
& = &  - \underbrace{\frac{KC}{J}}_{\mbox{{\small termo reativo}}}
+ \nonumber \\
\nonumber \\
+ D \underbrace{\frac{\pt}{\pt \xi} \left[ J \left( \alpha \frac{\pt C}{\pt \xi} - \beta \frac{\pt C}{\pt \eta} \right) \right.}_{\mbox{{\small termo difusivo}}} 
& + & D \underbrace{ \frac{\pt }{\pt \eta} \left. J  \left( \gamma \frac{\pt C}{\pt \eta} - \beta \frac{\pt C}{\pt \xi} \right) \right],}_{\mbox{{\small termo difusivo}}} 
\label{eq4}
\end{eqnarray}

\noindent onde $C= C(\xi, \eta, \tau)$ representa a concentração local de coliformes fecais ao longo do tempo. Os termos $K$ e $D$  são as 
constantes de decaimento e difusão dos coliformes fecais. Note que considera-se as difusões de coliformes fecais nas 
direções $\xi$ e $\eta$ iguais \cite{romeiro}. 

Os termos das equações (\ref{eq1}-\ref{eq3}) e (\ref{eq4}) estão agrupados de acordo com as  características (conservação de massa, temporal, 
convectivo, difusivo, pressão e reativo) que descrevem as propriedades do modelo. Note que os termos convectivos contém não-linearidades, que
em geral são tratadas com técnicas específicas \cite{fortuna, ferreira}.

As equações hidrodinâmicas (\ref{eq1}-\ref{eq3}) e a equação de concentração de coliformes fecais (\ref{eq4}) são discretizadas através do 
Método de Diferenças Finitas. O termo convectivo (não linear) é discretizado por meio do Método 
\textit{upwind }FOU (\textit{First order upwind}) \cite{maliska, fortuna}. Evidentemente existem técnicas mais sofisticadas e de maior 
acurácia do que o método FOU, no entanto para as condições iniciais e de contorno abordadas no presente trabalho o método FOU produz
resultados numéricos consistentes.

Para que o modelo analisado descreva a dinâmica no lago é necessário que as características reais do lago sejam preservadas no modelo. O domínio 
onde as equações (\ref{eq1}-\ref{eq4}) serão calculadas, também chamado de malha computacional, deve conter as características da geometria 
do lago Luruaco. O uso do método multibloco para regiões com contornos irregulares, como é o caso do lago Luruaco, preserva os contornos 
originais, podendo aumentar ou diminuir o grau de refinamento em regiões específicas da malha. Na próxima seção constrói-se a malha 
computacional do lago Luruaco.    

\newsec{Malha Computacional do Lago Luruaco}

A malha computacional é o domínio onde o modelo matemático-numérico deve ser simulado, de modo que quanto mais próxima da geometria real a 
malha for, mais realística será a simulação numérica.  

Nesse contexto, a fronteira do domínio do lago contém  informações à respeito das entradas e saídas de água do lago, além de caracterizar a 
geometria real do lago com a presença de curvas suaves e formas angulosas (bicos). Para captar essas características foi utilizado o 
programa   \textit{WebPlotDigitizer} \cite{rohatgi2011webplotdigitizer}, onde a partir de uma imagem ou gráfico são disponibilizadas as 
coordenadas dos pontos da região analisada. Foram coletadas 309 coordenadas do contorno do lago, os pontos foram 
interpolados através do método \textit{spline} cúbico parametrizado e o bordo foi obtido. Em seguida, via um conjunto de equações de geração
de malhas, o grid inscrito ao contorno foi obtido. Os detalhes sobre as técnicas matemáticas do processo de geração de malhas podem ser 
encontrados em \cite{cirilo, thompson}.

\begin{figure}[!ht]
	\centering
	\includegraphics[scale=0.62]{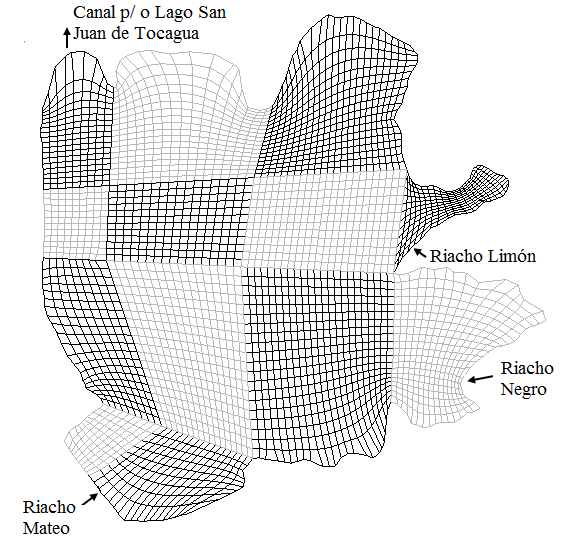}
	\caption{Malha multibloco do lago Luruaco composta por 13 sub-blocos.}
	\label{fig2}
\end{figure}

Quanto à geração do interior da malha computacional, utilizou-se o método multibloco. O método multibloco faz a construção da 
malha computacional através da união de sub-malhas menores, interligadas através de suas arestas. As equações do fluxo hidrodinâmico e de 
concentração de coliformes fecais são calculadas em cada sub-bloco e transmitidas para os sub-blocos adjacentes na forma de condição de contorno. Assim, 
garantida a comunicação entre os sub-blocos, a simulação numérica fornece a dinâmica de coliformes fecais em todo o lago Luruaco. 

Uma das vantagens para o uso do método multibloco está na questão do detalhamento da malha. Cada bloco pode ter suas linhas coordenadas ajustadas de maneira 
independente, desde que os blocos com adjacência sejam interligados sob o mesmo número de linhas. Essa construção permite que regiões que 
necessitem de maior detalhamento, como regiões irregulares ou entradas e saída de água, possam ser melhor adaptadas localmente. A Figura 2 
apresenta o domínio geométrico do lago Luruaco discretizado.


\newsec{Simulações Numéricas}

 Nesta seção apresenta-se as simulações numéricas do transporte de coliformes fecais no corpo d'água do lago Luruaco. Inicialmente ordenam-se 
 as características e as hipóteses utilizadas em nosso modelo a ser simulado numericamente.
 
 Sobre o lago Luruaco, o mesmo caracteriza-se por ser uma lâmina d'água, quando compara-se sua área com sua profundidade. Nesse contexto, um 
 modelo do tipo 2DH é utilizado nas simulações. Quanto à discretização da geometria do lago, é suposto que o lago Luruaco tem 3 fontes 
 (afluentes) e 1 sorvedouro (efluente), Figura 1. 
 
 Sobre o escoamento do corpo d'água, na modelagem é suposto que o transporte dos coliformes fecais ocorre de modo passivo. A hipótese do 
 escoamento passivo permite desacoplar a simulação numérica em duas partes, ou seja, primeiramente calcula-se o campo de velocidades 
 advectivo do escoamento da água na geometria modelada do lago, para posteriormente simular o transporte das concentrações de coliformes 
 fecais, nessa geometria, através do modelo advectivo-difusivo-reativo.
 
 Sobre o modelo hidrodinâmico, supomos que o escoamento da água é descrito pelas equações de Navier-Stokes e da pressão (\ref{eq1}-\ref{eq3}) 
 com as hipóteses do fluido ser incompressível do tipo newtoniano com coeficiente de Reynolds $Re=555$.
 
 Sobre o modelo de transporte de coliformes fecais (\ref{eq4}) considera-se que o campo de velocidades de escoamento é suposto ser a soma 
 vetorial dos campos advectivo, fornecido pelo modelo hidrodinâmico, e do campo difusivo. Supõem-se também que os coeficientes de difusão 
 molecular $D$ e de decaimento $K$ dos coliformes fecais sejam constantes em toda a geometria do lago. 
 
 Sobre o coeficiente de difusão $D$, note 
 que a difusão molecular localmente espalha os coliformes fecais por movimento aleatório, mas, em larga escala, são os redemoinhos (vórtices) 
 e turbilhões que espalham os coliformes fecais através da chamada difusão turbulenta. Como a escala dos redemoinhos são muito maiores que 
 as escalas da difusão molecular, a difusão turbulenta é várias ordens de grandeza maior que aquela da difusão molecular. Conforme 
 Chapra \cite{chapra}, as várias espécies reativas têm valores de difusão molecular no intervalo de valores 
 entre $D=10^{-3}\;m^2/h$ e $D=10^{-1}\;m^2/h$. Já o coeficiente de difusão turbulenta, em lagos, que depende da escala do fenômeno 
 turbulento, assume valores entre $D=10^{1}\;m^2/h$ e $D=10^{10}\;m^2/h$. 
 
 Com o objetivo de simular em nossa modelagem os fenômenos de difusão 
 molecular e turbulenta, envolvidos no transporte de coliformes fecais na lâmina d'água do lago Luruaco, e devido à semelhança dos escoamentos 
 no lago Luruaco e no Lago Igapó I, localizado no município de Londrina, Paraná, supomos para ambos os mesmos valores para $D$. 
 Em \cite{Suellen} o melhor ajuste para o coeficiente de difusão foi $D=3,6\;m^2/h$. Quanto ao coeficiente de decaimento de coliformes 
 fecais, em \cite{romeiro} o melhor ajuste foi $K= 0,02 h^{-1}$, que será utilizado nessa simulação.
 
 Neste contexto, utilizando o nosso modelo matemático, vamos descrever qualitativamente o impacto que uma descarga contínua de coliformes 
 fecais, nas 3 entradas (afluentes) do lago Luruaco, produz em toda sua extensão. Considera-se as seguintes condições iniciais e de fronteira.
 
 \vskip 0.35cm
 {\bf{Condições iniciais para o modelo hidrodinâmico}}
 \vskip 0.35cm
 
 \noindent 
 Considerou-se no instante inicial, para os campos de velocidades e de pressões, que o lago encontra-se em estado de quiescência, ou seja,
 suas águas estão supostamente paradas em todo o domínio, exceto nas regiões dos afluentes e efluentes. Nos afluentes 1, 2 e 3 foram consideradas
 entradas instantâneas de água, e saída instantânea de água pelo efluente, de modo que o coeficiente de Reynolds adotado ($Re=555$) preserve conservação da massa.
 
\vskip 0.35cm
{\bf {Condições de fronteira para o modelo hidrodinâmico}}
 \vskip 0.35cm
 
 \noindent
 Considerou-se que o campo de velocidades, para $t>0$, é nulo em toda a fronteira da geometria do lago, exceto nas entradas (afluentes) e saída (efluente) onde considerou-se os mesmos valores da condição inicial. Quanto à pressão, a condição tipo Neumann foi aplicada em toda a borda da geometria.

\vskip 0.35cm
{\bf{Condições iniciais para o modelo de transporte e reações}} 
\vskip 0.35cm
\noindent
	Considerou-se que no instante inicial o campo escalar de concentrações de coliformes fecais é nulo em todos os ponto da malha (interiores e de fronteira) do lago. 
	
\vskip 0.35cm	
{\bf{Condições de fronteira para o modelo de transporte e reações}}
\vskip 0.35cm

\noindent	
	Para $t>0$ considerou-se as seguintes concentrações constantes nas entradas e saídas do lago Luruaco: 
	\begin{eqnarray}
	&&C({X}_{Negro},t)=100\;MPN/100 ml, \nonumber \\
	&&C({X}_{Mateo},t)=100\;MPN/100 ml \nonumber \\
	&&C({X}_{Limon},t)=500\;MPN/100 ml \label{eq49}\\
	&&C({X}_{efluent},t) \quad \textrm{condição tipo Neumann}, \nonumber
	\end{eqnarray}
	\noindent 
	onde utilizamos a seguinte notação compacta ${X}_{Negro}=(x,y)_{Negro}$, as coordenadas dos pontos de fronteira da malha que pertencem à entrada do Riacho Negro, idem para os demais afluentes e efluentes do lago. A  unidade $MPN/100$ ml significa {\it 
	Most Probable Number} (Número Mais Provável) de coliformes fecais por 100 ml em uma amostra de água. Nos demais pontos da fronteira toma-se condição tipo Neumann.
		
	A seguir são apresentadas as simulações numéricas para o campo de velocidades, Figura 3, e para o campo de concentrações de coliforme
	fecais, Figura 4. As simulações foram realizada em um período de 72 horas, até que o regime estacionário fosse atingido.
	Salienta-se que a Figura 3, à esquerda, exibe a magnitude do vetor resultante da velocidade, enquanto que a Figura 3, à direita,
	mostra a direção e sentido do vetor resultante da velocidade.
	
	\begin{figure}[!htb]
	\begin{center}
	\subfigure{\includegraphics[scale=0.31]{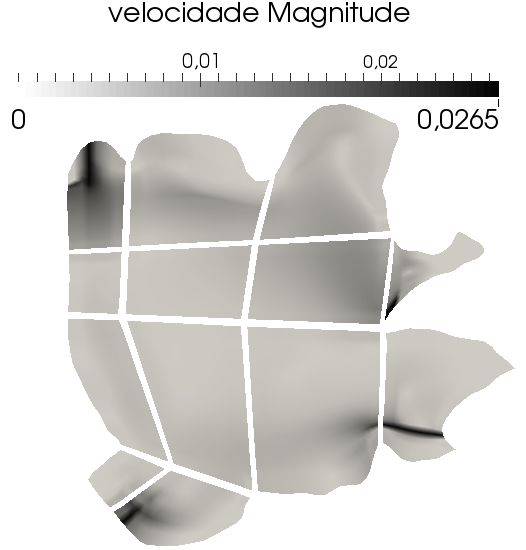}} \qquad
	\subfigure{\includegraphics[scale=0.38]{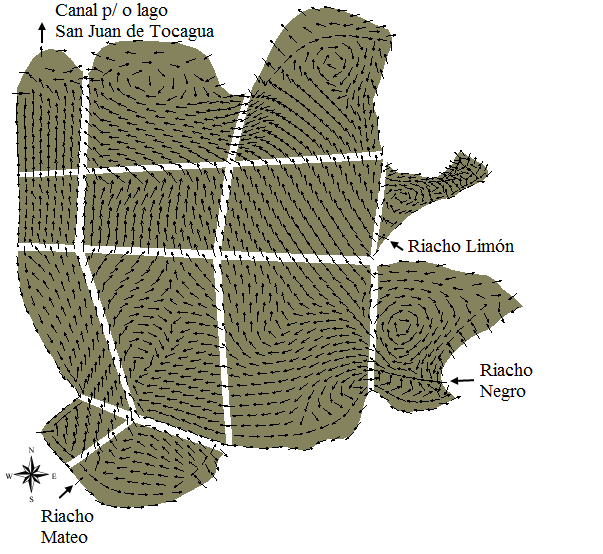}}
	\caption{Simulação numérica do campo de velocidades gerado na malha multibloco do lago Luruaco para $Re=555$. 
	 Mapa com gradiente de tons de cinza (esquerda) e mapa de campo vetorial (direita).}
	\end{center}
	\label{fig3}
	\end{figure}

\begin{figure}[!htb]
	\centering
	\subfigure{\includegraphics[scale=0.83]{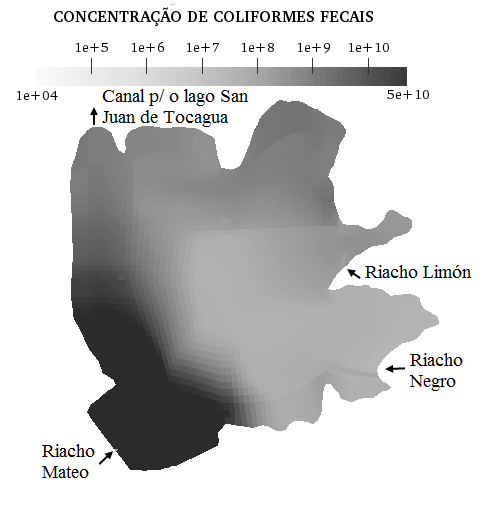}}
	\caption{Simulação numérica do campo de concentrações de coliformes fecais gerado na malha multibloco do lago Luruaco para $D=3,6\;m^2/h$ e  $K= 0,02 h^{-1}$.}
	\label{fig4}
\end{figure}

 \newsec{Considerações Finais}
 
Através das simulações numéricas do sistema de equações (\ref{eq1}-\ref{eq3}) e (\ref{eq4}) pretende-se descrever qualitativamente 
a dinâmica do transporte de coliformes fecais no corpo d'água do lago Luruaco. Salienta-se que essa simulação numérica não tem a pretensão 
de fornecer previsões quantitativas sobre o índice de poluição em um dado local do domínio físico do lago, num dado instante. Sabe-se que 
as condições de entrada (iniciais e fronteira) de coliformes fecais variam diariamente. Nessas condições, o nosso objetivo é fornecer 
informações qualitativas, por exemplo, os locais mais poluídos no domínio do lago, independentemente das concentrações iniciais e de fronteira. Assim, estudar a dinâmica de coliformes fecais no corpo d'água do lago Luruaco é uma forma de determinar as regiões que apresentam maior risco
de contaminação para a população local.

Observando a Figura 4, nota-se que a escala de concentrações de coliformes fecais varia de $C=10^{4} MPN/m^3=1 MPN/100ml$, correspondente a excelente qualidade de água, até  $C=5 \times 10^{10} MPN/m^3=5 \times 10^{6} MPN/100ml$, que corresponde a águas muito poluídas. 

Como consequência da geometria do lago Luruaco e do campo hidrodinâmico de velocidades, Figura 3, nota-se que a poluição injetada no lago 
Luruaco pelo Riacho Negro flui em parte na direção do Riacho Mateo, somando-se à poluição injetada por esse último, gerando nessa região 
concentrações altas de coliformes fecais, Figura 4. Na margem oposta, margem norte, devido à injeção de coliformes pelo Riacho Limón, tem-se 
também uma região muito poluída que se estende até o Canal efluente para o Lago San Juan de Tocagua. Nota-se que  as regiões central e 
sudeste do lago apresentam a melhor qualidade de água com uma concentração de coliformes fecais relativamente baixa, Figura 4.

Enfim, essas simulações numéricas permitiram uma melhor compreensão qualitativa da dinâmica de coliformes fecais no lago Luruaco, evidenciando as regiões mais poluídas. Assim, o uso da simulação numérica é um instrumento útil e importante para avaliar a qualidade da água em massas de água, apresentando resultados que podem ser adotados por órgãos públicos para a recuperação do ambiente poluído, para a identificação de emissores  de poluentes e para proporcionar uma melhor qualidade de vida para os utilizadores do lago, bem como para os habitantes da cidade de Luruaco que dependem dessa água.

\vspace{0.7cm}
\noindent{\bf\Large{Agradecimentos}}
\vspace{0.5cm}

\noindent Os autores agradecem o suporte financeiro da CAPES, oriundo do edital 24/2012 - Pró-Equipamentos Institucional, convênio 774568/2012.

\begin{abstract}
	
{\bf Abstract}
	
The Luruaco Lake located in the Department of Atlántico, Colombia, is damaged by the discharge of untreated sewage, bringing risks to the health of all who use its waters. The present study aims to perform the numerical simulation of the concentration dynamics of fecal coliforms in the lake. The simulation of the hydrodynamic flow is carried out by means of a two-dimensional horizontal (2DH) model, given by a Navier-Stokes system. The simulation of fecal coliform transport is described by a convective-dispersive-reactive equation. These equations are solved numerically by the Finite Difference Method (FDM) and the Mark and Cell (MAC) method, in generalized coordinates. Regarding the construction of the computational mesh of the Luruaco Lake, the cubic spline and multiblock methods were used. The results obtained in the simulations allow a better understanding of the dynamics of fecal coliforms in the Luruaco Lake, showing the more polluted regions. They can also advise public agencies on identifying the emitters of pollutants in the lake and on developing an optimal treatment for the recovery of the polluted environment.
	
{\bf Keywords}. Luruaco Lake, Fecal Coliforms, Multiblock Mesh, Finite Difference Method, Mark and Cell Method, Curvilinear Coordenates.
\end{abstract}

\end{document}